\RequirePackage{graphicx} 
\documentclass[%
 aip,
 amsmath,amssymb,
 reprint,%
]{revtex4-1}

\usepackage{graphicx}
\usepackage{dcolumn}
\usepackage{bm}
\usepackage{hyperref}
\usepackage[utf8]{inputenc}
\usepackage[T1]{fontenc}
\usepackage{mathptmx}
\usepackage{amsmath}
\usepackage{multirow}
\usepackage{makecell}
\usepackage[inline, nomargin]{fixme}
\usepackage{physics}

\graphicspath{{Pictures/}}
\begin{document}

\preprint{AIP/123-QED}

\title[Automated analysis of single-tone spectroscopic data for cQED systems]{Automated analysis of single-tone spectroscopic data for cQED systems\\~}

\author{G. P. Fedorov}
\email{gleb.fedorov@phystech.edu}

\affiliation{%
National University of Science and Technology MISiS, Moscow, Russia
}%
\affiliation{ 
Russian Quantum Center, Skolkovo, Moscow, Russia 
}%
\affiliation{ 
Moscow Institute of Physics and Technology, Dolgoprudny, Russia
}%

\author{A. V. Ustinov}
\affiliation{%
National University of Science and Technology MISiS, Moscow, Russia
}%
\affiliation{ 
Russian Quantum Center, Skolkovo, Moscow, Russia 
}%
\affiliation{%
Karlsruhe Institute of Technology, Karlsruhe, Germany
}%

\date{\today}

\begin{abstract}
Physical systems for quantum computation require calibration of the control parameters based on their physical characteristics by performing a chain of experiments that gather most precise information about the given device. It follows that there is a need for automated data acquisition and interpretation. In this work, we have developed a tool that allows for automatic analysis of single-tone spectroscopy (STS) results for a single cell consisting of qubit and resonator in the circuit quantum electrodynamics (cQED) architecture. Using analytic approaches and maximum likelihood estimation, our algorithm is capable of finding all relevant physical characteristics of the cell by using only the measured STS data. The described approach is fast and robust to noise, and its open-source Python implementation can readily be used to calibrate transmon qubits coupled to notch-port resonators.
\end{abstract}

\maketitle

 \renewcommand*{\figureautorefname}{Fig.}

\section{Introduction} \label{sec:level1} 

Computation on a quantum computer involves operating large numbers of physical quantum bits (qubits). One of the most significant challenges here is that individual man-made qubits are not identical and cannot all be operated same way\cite{kelly2018, chen2018}. Typically, control parameters for a device depend on its physical parameters and are determined by a sequence of calibration steps. Since the devices cannot be produced identical, the optimal parameters vary among them and should be found \textit{in situ}. Manual adjustment of these parameters is not a scalable solution, and operating just a handful of qubits already requires a computer system to handle the calibration on its own. Moreover, a lot of research is still being done to optimize chip designs and fabrication methods. This research includes gathering statistics on many different samples that should be measured in a comparable way and consistently evaluated. An automated measurement system does not only speed up this process, but also excludes human errors.

In this work, we are proposing and implementing several methods of data processing and computer vision that aid automatic calibration of circuit quantum electrodynamics (cQED)\cite{blais2007} cells used to read out the quantum states of superconducting qubits. Computer vision is understood here as an enterprise that employs statistical methods to disentangle data using models constructed with the aid of geometry, physics, and learning theory.\cite{forsyth2011} As it will be shown below, the outputs of circuit calibration on certain steps can to be represented as images, and it is hardly possible to automatically extract the required information from them without resorting to methods of computer vision. The approach that we propose is accurate, fast and robust to noise, it is applicable to superconducting qubits with a spectrum periodic in magnetic (or electric) field. The generic single cell here is described with closed-form model equations, and is compatible with the paradigm of one readout resonator per one qubit.\cite{versluis2017, kelly2015} 

\begin{figure*}
	\centering
	\includegraphics[width=0.95\linewidth]{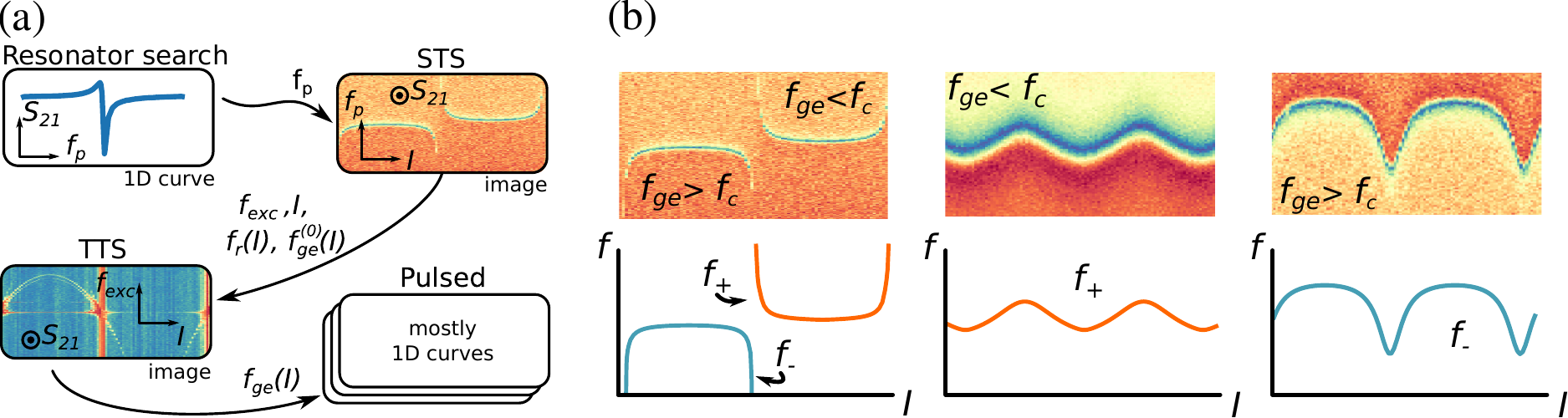}
	\caption{(a) Typical experiment flow to calibrate a tunable-frequency qubit. Usually, the single- and two-tone spectroscopic measurements are performed before starting pulsed experiments, and each next measurement is based on the results of the previous. (b) Examples of the expected outputs (lower row) for three types of qualitatively different STS data (upper row). The 2D heatmaps should be brought into correspondence with 1D model curves. In the first column, the so called avoided crossing pattern is shown where two model curves $f_+(I)$ and $f_-(I)$ have to be fitted simultaneously since the model is discontinuous (the qubit transition frequency $f_{ge}(I)$ may be tuned both above or below the bare resonator frequency $f_c$). Conversely, the second and the third heatmaps should be described by a single continuous curve: it is $f_+(I)$ for the former (where $f_{ge}(I)<f_c;\ \forall I$) and $f_-(I)$ for the latter ($f_{ge}(I)>f_c;\ \forall I$).}
	\label{fig:detection}	
\end{figure*}

Within the considered paradigm, the flow of an experiment that characterizes a single qubit-resonator cell of a quantum processor chip is depicted schematically in \autoref{fig:detection}(a). It is similar to the  procedures for measuring cQED samples of qubits with tunable frequency described in Refs.[\onlinecite{jerger2013, chen2018}]. For non-tunable qubits, a different algorithm of automatic calibration was described in the patent application Ref. [\onlinecite{bloom2018}]. 

We start from briefly discussing each step shown in \autoref{fig:detection}(a). Usually, one starts by locating the dip in the transmission corresponding to the resonance frequency of the readout resonator. Having found the dip, the experimenter can set the range to scan the probe frequency $f_p$. Scanning this range during the next step produces single-tone spectroscopy (STS) data. STS tracks the frequency shift of the resonance dip while changing the control parameter (e.g., the magnetic field applied to the sample that is proportional to some control current $I$). By processing the results of the STS, one can get a coarse estimate on the qubit frequency dependence on current $I$  denoted as  $f^{(0)}_{ge}(I)$ (here $g$ and $e$ stand for ground and excited state of the qubit, respectively). Besides that, during this step one also gets accurate estimates of the resonator frequency $f_r(I)$ and the resonator-qubit coupling strength $g$. Based on these parameters, one can later set the scan ranges for the excitation frequency and the current ($f_{exc}$ and $I$) to perform the two-tone spectroscopy (TTS). Finally, by processing the TTS results, one can obtain the precise dependence of the qubit transition frequency $ge$ on current $f_{ge}(I)$	 with accuracy sufficient for pulsed experiments in order to perform accurate single-qubit gates. Overall, the process can be structured so that the results of one measurement step define the parameters of the next, successively acquiring more accurate information about the physical system.

As one can see, the results that are obtained in this scheme may be divided in two groups: some contain single curves (data are 1D manifolds), and others contain colour scale images (heatmaps, 2D manifolds). The work presented in this paper is confined to the automatic analysis of the latter. More precisely, we are interested in processing STS data. In \autoref{fig:detection}(b) we illustrate three types of correlated measured data and the theoretical model. Depending on the range of the qubit transition frequencies relative to the resonator frequency, we need to interpret STS data as either a single curve or two curves defined by the cQED model. Here we are aiming at building a method that is able to distinguish between these three cases and to estimate quantitatively the physical parameters of the system based on the best-fit model. 

The model to fit the data is based on the theoretical description of the cQED systems and physical qubits. In Appendices \ref{sec:transmon} and \ref{sec:cqed} we derive all the necessary equations to form the model curves that are expected to appear in single-tone spectroscopy. We use a transmon\cite{koch2007} with an asymmetric SQUID (superconducting quantum interference device) as a qubit. Since our methods do not depend on the particular shape of the magnetic (or electric) field dependence of the qubit energy levels, without any loss of generality the presented approach can also be used for other types of qubits.

For the singe-tone spectroscopy, the theoretical curves are described by the function (see \autoref{fig:detection}(b)):
\begin{align}
f_\pm(I) \equiv f_r(I) = \frac{f_c + f_{ge}(I)}{2} \pm \sqrt{g^2+(f_{ge}(I) - f_c)^2/4},\label{eq:f_r}
\end{align}
where $f_c$ stands for the field independent resonator frequency; $f_r(I) = f_c$ when $g=0$ (it is also defined as the \textit{bare cavity frequency}, or the frequency of the cavity with no qubit coupled to it). Next, the qubit transition $f_{ge}(I)$ is expressed as
\begin{equation}
f_{ge}(I) = f_{ge}^{max} \left[\cos^2\left(\frac{\pi(I-I_{ss})}{\Pi}\right)+d^2 \sin^2 \left(\frac{\pi(I-I_{ss})}{\Pi}\right)\right]^\frac{1}{4},
\label{eq:tr_spectrum}
\end{equation}
where $\Pi$ is the period of the spectrum in current $I$, $d$ is the SQUID asymmetry parameter and $f_{ge}^{max}$ is the qubit frequency at the ``sweet spot'' corresponding to current $I = I_{ss}$, where the applied magnetic field exactly compensates the non-zero local stray field that is usually present on the chip in a real experiment.

We note that for the avoided crossings pattern both curves from \eqref{eq:f_r} are necessary. Alternatively, for the cases when the qubit frequency is entirely below (above) the bare cavity frequency only $ f_+(I)\ \left(f_-(I)\right)$ is used. 
\begin{table}
\begin{ruledtabular}
	\begin{tabular}{rl} 
		Parameter & Physical meaning \\ 
		\hline
		$f_c$, GHz & bare cavity frequency \\ 
		$g$, GHz & cavity-qubit coupling constant \\
		$\Pi$, A & period in current of the transmon \\
		$I_{ss}$, A & transmon upper sweet-spot current \\
		$f_{ge}^{max}$, GHz & transmon frequency at sweet-spot \\
		$d$& transmon asymmetry parameter
	\end{tabular} 
\end{ruledtabular}
\caption{Parameters of the model and their physical meaning.}
\label{tab:pars}
\end{table}
Overall, we thus have six parameters for our model: $f_c$, $g$, $\Pi$, $I_{ss}$, $f_{ge}^{max}$ and $d$. Since they are the parameters in the Hamiltonian for the compound cavity-qubit system described in Appendix \ref{sec:transmon}, hereafter we call them the Hamiltonian parameters.

The structure of the rest of the paper is as follows. First, we describe our approach to extract the Hamiltonian parameters from the STS data. Then we address issues related to the accuracy, performance, and reliability of the algorithm. Finally, we discuss the limitations of the algorithm, further applications and future work. 

\section{Methods}

Before describing the methods, we first highlight important peculiarities of the data and the essential experimental details. A sample in the cQED architecture interacts with incident microwave radiation. Therefore, the standard mathematical concept to describe its properties is the scattering matrix $S_{ij}$ which shows how harmonic signals scatter between $i$\textsuperscript{th} and $j$\textsuperscript{th} ports of the sample. The absolute value of the complex scattering parameter $S_{21}(f_p)$ shows the amplitude of the signal transmission between the ports 1 and 2 at some probe frequency $f_p$. Alternatively, $\angle S_{11}(f_p)$ would show the change of the signal phase after the reflection from the port 1. By varying the probe frequency and other control parameters, one can record changes in $S_{ij}$ and from that draw conclusions about the properties of the studied device.

\begin{figure}
\includegraphics[width=\linewidth]{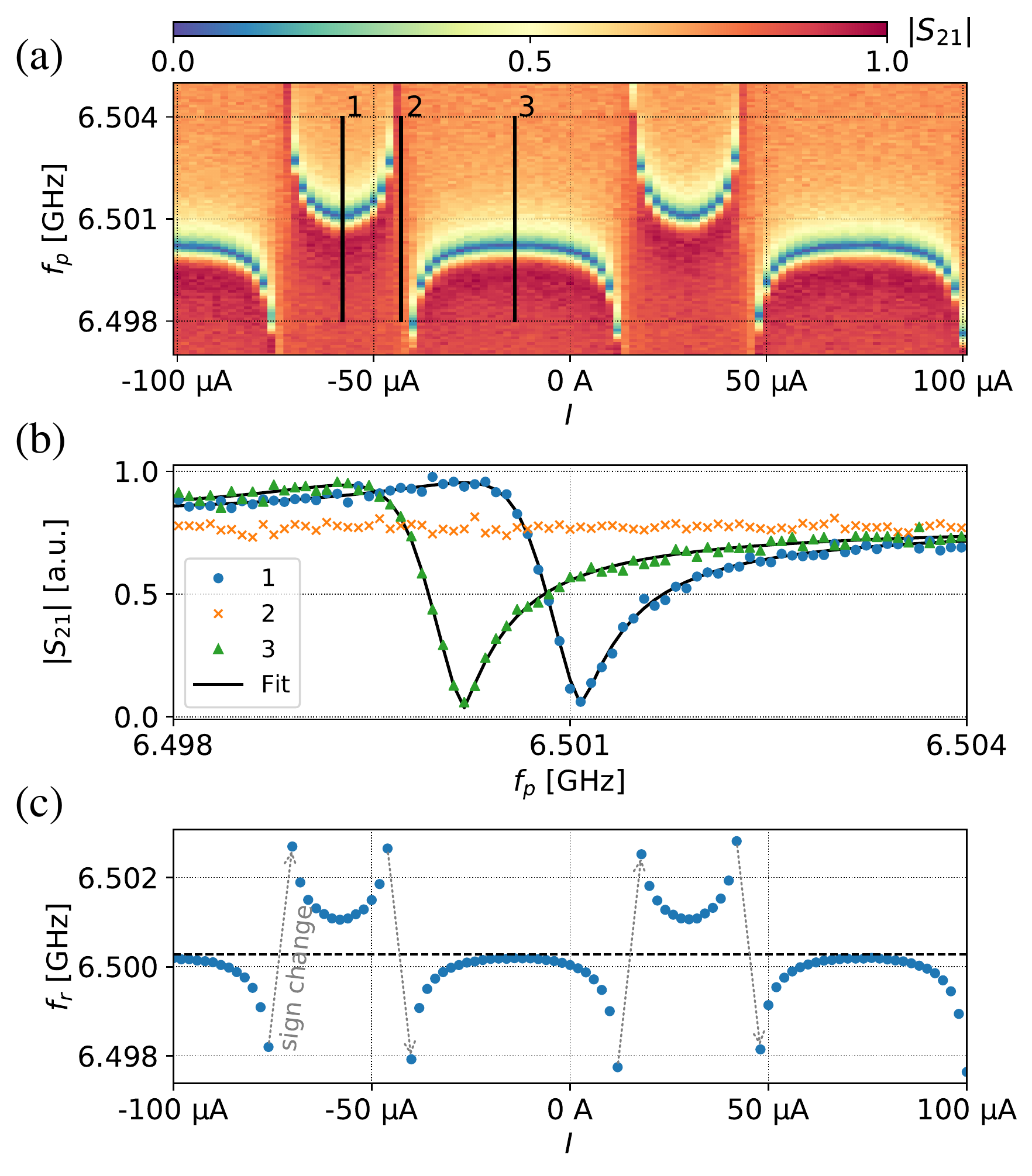}
\caption{(a) Raw experimental heatmap of $|S_{21}|$ depending on the probe frequency $f_p$ and the coil current $I$. (b) Slices of the transmission from (a) showing two slices (1,3) with fits and a plateau 2 with no dip. (c) Extracted resonance frequency $f_r(I)$ (blue dots) and mean frequency value over all currents $\langle f_r \rangle_{I}$ (black dashed line). Grey arrows show where $f_r - \langle f_r \rangle_{I}$ changes sign.}
\label{fig:anti_exp}
\end{figure}

As an example, in \autoref{fig:anti_exp}(a) we show a heatmap for the amplitude $|S_{21}|$ obtained in STS for a notch-type (side-coupled) coplanar waveguide resonator connected to a tunable transmon qubit. $|S_{21}|$ is shown in colour, and two other axes show the probe frequencies $f_p$ and the coil current $I$ values for which it was recorded. Looking at the data slices in \autoref{fig:anti_exp}(b), one can say that the plot shows the dependence of the resonator dip frequency on the applied coil current. 

While in \autoref{fig:anti_exp}(a) we have shown only the amplitude $|S_{21}|$,  the phase of the signal $\angle S_{21}$ is also important. The amplitude has a local minimum at resonance, while the phase has a local maximum of its derivative there. For certain kinds of readout resonators the phase provides better contrast than the amplitude and should be preferred. This means that, in general, the interpretation of the data requires more advanced analysis than simply identifying the location of the minimal transmission amplitude.

Now, knowing the internal structure of the data, we outline our algorithm of extracting the Hamiltonian parameters from them. The main idea is to fit the model \eqref{eq:f_r} to the data in \autoref{fig:anti_exp}(a) in the sense of maximum likelihood estimation\cite{bishop2006} (MLE, see also Appendix \ref{sec:ML}). To reduce the complexity of the estimation in terms of the number of parameters, we first perform the MLE of the resonator frequency by fitting only along the $f_p$ axis using an external fitting library called \textit{circlefit}\cite{probst2015} (see Section \ref{sec:extract_fr} for details). This removes the extra dimension from the data which now can be compared directly with the model leaving only 6 fit parameters. With an assumption\footnote{We note that the normality of noise in estimated $f_r$ is asymptotic\cite{jennrich1969} and would only be attained if the number of data points during the resonator fit was infinite. Therefore, this assumption is in fact an approximation\cite{anastasiou2017}. It is not essential for our method to be statistically consistent,\cite{jennrich1969} but allows to explicitly find the likelihood function and calculate the Cramer-Rao lower bounds for the fit parameters variances.}  that the distribution of noise in the estimated $f_r$ is normal, the MLE is equivalent to a nonlinear least-squares problem with global optimization. It is a complex task, firstly, due to the periodic dependence of the frequencies on current $I$ and the unknown position of the qubit sweet spot $I_{ss}$ and, secondly, due to the strong nonlinear dependence of Eqs. \eqref{eq:f_r} and \eqref{eq:tr_spectrum} on other parameters. Fortunately, it is possible to get a good initial guesses for $\Pi$ and $I_{ss}$ and then do a brute force search for the solution in other variables due to the well-defined bounds for each of them. Finally, we can fine-tune the result in all six parameters using a local optimizer.

The detailed description of the method which uses \autoref{fig:anti_exp}(a) as an example is presented below. As it has already been mentioned, the chosen type of the resonator-qubit arrangement which yields the avoided crossings pattern is not unique; however, the other two cases are treated the same way.

\subsection{Extracting resonance frequency}\label{sec:extract_fr}

The resonator curve MLE is based on the \textit{circlefit} library which does the least-squares fitting of various types of microwave resonators. For each current $I$, it fits the complex transmission $S_{21}(f_p)$ which, for a single ideal resonator, is a circle on the complex plane. However, when an real experimental device is considered, the response model becomes more complex. For example, for a notch-type resonator used in this work the following model is employed:\cite{probst2015}
\begin{equation}
S_{21}^{notch}(f_p) = ae^{i\alpha}e^{2\pi if_p\tau} \left[1-\frac{Q_l/Q_e'}{1+2iQ_l(f_p/f_r-1)}\right],
\label{eq:res_S21_probst}
\end{equation} 
where $a$ denotes the overall attenuation (amplification) in the transmission lines connecting the resonator to the measurement apparatus, $\alpha$ is the frequency-independent phase shift in the transmission lines, $\tau$ is the propagation delay far from the resonance; $Q_l$, $Q_e'$ are the usual loaded and complex\cite{khalil2012} external quality factors, respectively; $f_r$ is the resonance frequency. Fitting procedure of the $circlefit$ library is to sequentially extract all of the model parameters by a succession of partial approximations (see \cite{probst2015} for details). For brevity reasons, we do not present here the full complex fits for our data; however, as an illustration in \autoref{fig:anti_exp}(b) one can see the absolute values of the optimized models (solid black lines) that match well with the absolute values of the experimental data taken from slices 1 and 3 of \autoref{fig:anti_exp}(a). Finally, from the optimal model parameters we obtain the resonance frequency $f_r(I)$ depending on current. A possible caveat is that for some $I$ values the resonance dip may disappear so the resonator fit would fail (this happens, e.g., at the avoided level crossings; see \autoref{fig:anti_exp}(b), slice 2). Therefore, such slices are excluded in advance via a threshold condition.

The resulting plot of the extracted resonator frequency versus current $I$ is shown in \autoref{fig:anti_exp}(c) (blue dots). There are some current values located between the branches where the data points are missing, as expected, due to the absence of the resonance. Additionally, we plot here the mean value of the detected frequencies shown as a dashed black line. This parameter is important since it will, firstly, serve as an initial guess for the cavity frequency of the model \eqref{eq:f_r} with $f_c \approx \langle f_r \rangle_{I}$ and, secondly, will be used in the period and phase extraction algorithm which tracks the changes of the sign of the value $\Delta f_r = f_r - \langle f_r \rangle_{I}$ (marked as grey dashed lines in \autoref{fig:anti_exp}(c)). Locating these sign changes allows one to find the qubit sweet spot without fitting the full model.

\subsection{Extracting field period and sweet spot locations}

As we already have stated above, one of the obstacles for the fitting is the periodicity of the data in one of the fitting parameters. Along with the equally unknown phase of the signal, this leads to the presence of many local minima in the loss function which impede the progress of iterative optimization algorithms. In other words, the unknown parameters $\Pi$ and $I_{ss}$ are preventing us from finding the global minimum. So it would be very convenient to determine the period and the sweet spot location without fitting the full model \eqref{eq:f_r} to alleviate the stated problem. In the absence of noise, this would be a trivial task. Indeed, one would just need to find the sign changes in $ \Delta f_r $ directly by traversing the array of its values, and then compute the period as the distance between two similar sign changes, and the sweet spot location as the point between the adjacent ones. However, in the conditions of experiment, the noise to some extent is always present, therefore below we suggest a solution that is insensitive to local sporadic perturbations of the data.

\begin{figure}
	\centering
	\includegraphics[width=\linewidth]{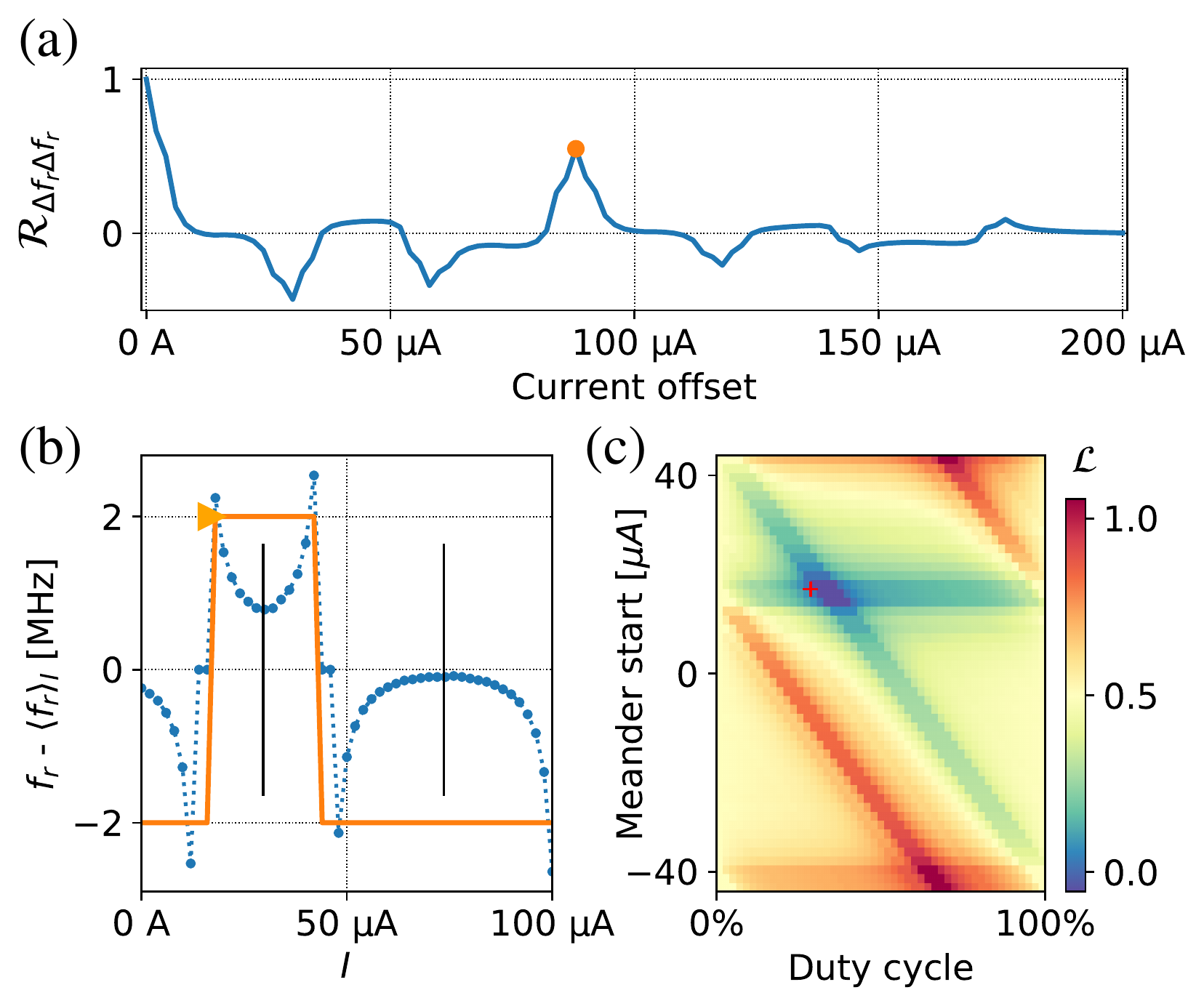}
	\caption{Period and phase extraction procedure. (a) Autocorrelation function depending on the current offset $\Delta I$ shows a prominent local maximum at $88\ \mu$A (orange dot). (b) Construction for phase estimation; $\Delta f_r = f_r-\langle f_r \rangle_{I}$ (blue) is fitted with a square wave (orange, start marked with a triangle). Vertical bars mark candidate sweet spots. (c) Loss function (normalized) for the square wave fitting procedure from (b); red cross indicates the parameters of the meander shown there.}
	\label{fig:per+phase}
\end{figure}

An obvious method to find analytically the period and the phase (here, the phase is equivalent to $I_{ss}$) of a signal is the Fourier transform. Fast Fourier transform (FFT) should be applied to the data, and then the peak of the largest magnitude in the spectrum would give the likely frequency and the phase of the signal. FFT works well for the continuous curves, but it becomes inconvenient for the avoided crossings pattern due to the complexity of its spectrum caused by the discontinuity. Moreover, FFT may be not sufficiently accurate when the data contains only a couple of periods. A more robust tool to find the period in a given dataset $y$, especially when it contains just a few periods, is the autocorrelation function $\mathcal{R}_{y y}(l) = \sum_n y_n y_{n-l}$. The location of the largest of its local maxima (except for $l=0$) equals exactly the sought period\cite{parthasarathy2006}. However, this is only true when the mean value of the function is zero; otherwise, $\mathcal{R}_{y y}(l)$ would be linear with a steep slope that may smear out all the extrema. Therefore, instead of directly calculating the autocorrelation of $f_r (I)$ we consider the standardized function $\Delta f_r (I)$ introduced above which has zero mean and the same period. For this function, the autocorrelation function $\mathcal{R}_{\Delta f_r \Delta f_r}(\Delta I)$ depending on the current offset $\Delta I$ is shown in \autoref{fig:per+phase}(a). As one can see, $\Delta I$ spans 200 $\mu$A just as the data itself. This means that to calculate $\mathcal{R}_{\Delta f_r \Delta f_r}(\Delta I)$ the data is being zero-padded at all $\Delta I$ except for $\Delta I = 0$, and this is why we get diminishing correlation peaks at $\Pi$, $2\Pi$, etc. The orange dot in the plot \autoref{fig:per+phase}(a) shows the highest local extremum of $\mathcal{R}_{\Delta f_r \Delta f_r}(\Delta I)$ at 88 $\mu$A. It is a very prominent peak and can be easily distinguished among all others. There is as well a small peak at 176 $\mu$A which corresponds to $\Delta I = 2\Pi$. Note that the autocorrelation function is not ideally smooth and has some abrupt bends on the sides of the peaks. This happens because of some missing points in the resonator frequency data (corresponding to the plateaus of \autoref{fig:anti_exp}(b)) that were added back as zeros since the step between adjacent data points should be identical for the autocorrelation method to be accurate. This zero padding may be observed directly in \autoref{fig:per+phase}(b).

The problem with the autocorrelation function is that it does not tell us anything about the phase of the signal, i.e. the exact value of the sweet spot current. Therefore, after having found the period, we need another method to precisely determine $I_{ss}$. This method consists of finding the global maximum of the zero-lag correlation function $\mathcal{R}_{\Delta f_r S}(0)$ between $\Delta f_r(I)$ and a square wave $S(I, \Pi, \phi, D)$ having the same period but unknown phase $\phi$ and duty cycle $D$ which is the relation between the horizontal spans of the ``high'' and ``low'' levels. 
The values of the ``high'' and ``low'' levels must be opposite, i.e. 1 and -1, while their magnitude does not matter.  An illustration of a square wave function satisfying the optimal condition is presented in \autoref{fig:per+phase}(b) in orange. The phase $\phi$ (in $\mu$A) denotes the $I$-coordinate of the first point after the rising edge, and is marked with a triangle. Generally, the idea behind this approach is to robustly detect sign changes that were shown back in \autoref{fig:anti_exp}(c). Since $\mathcal{R}_{\Delta f_r S}(0) = \sum_I \Delta f_r(I) \cdot S(I) $, it gains value if $\Delta f_r$ and $S$ are of the same sign for a certain $I$. Therefore, the combination of $\phi, D$ that ensures the maximal number of points where $f_r(I)\cdot S(I)>0$ (the one that we are looking for) delivers the global maximum to $\mathcal{R}_{\Delta f_r S}(0)$. This approach will still work correctly even if the mean value $\langle f_r \rangle_{I}$ does not lie exactly between the branches in the avoided crossings pattern and intersects one of them since the period $\Pi$ is already found and fixed. 

The optimal values for $\phi$ and $D$ are found using the brute force algorithm. It calculates the loss function $\mathcal{L} = - \mathcal{R}_{\Delta f_r S}(0)$ on a $50 \times 50$ grid of $(\phi, D)$ and takes the minimal value of all. This method is stable and universal due to the evident boundaries on $\phi \in [-\Pi/2,\Pi/2)$ and $D \in [0, 1]$. The loss function topography for the avoided crossing patterns is nicely structured and for our example is shown in \autoref{fig:per+phase}(c). One peculiarity is that instead of a single minimum it has an area of the same minimal value. Again, this effect comes from the missing zero-padded $f_r$ points at some $I$ values. However, any value from this valley suits well enough for our purposes, and the algorithm finds no difficulty in locating it.

Having found values for $\Pi,\ \phi$ and $D$ we now can finally calculate the current of the transmon sweet spot in the case of the avoided crossings pattern:
\begin{align*}
I_{ss} = 
 \phi + \Pi (1+D)/2.
\end{align*}
Alternatively, for the continuous patterns (which are expected for $f_{ge}(I)<f_c$ or  $f_{ge}(I)>f_c$) we need to apply a different formula:
\begin{equation*}
I_{ss} = \phi + \Pi D/2.
\end{equation*}
Since at this point the algorithm does not yet know which type of the pattern it observes, it may apply some heuristic criterion to guess which equation to apply. When the noise is not too large, such heuristic can be to calculate the maximal absolute differential of the frequency data: $\max_{i>0} |f_{r,i} - f_{{r,i}-1}|$, and compare it to the peak-to-peak amplitude: $\max_{i,j} | f_{r,i} - f_{r, j}|$. For the avoided crossings these values are close and for the smooth dependencies they are not. However, in case of a strong noise this indicator may fail, and the program will have to check both current values to be $I_{ss}$ by fitting the full model two times, and then choose between the two possibilities based on the maximal likelihood.

\subsection{Full model fitting}

\begin{figure}
	\centering
	\includegraphics[width=\linewidth]{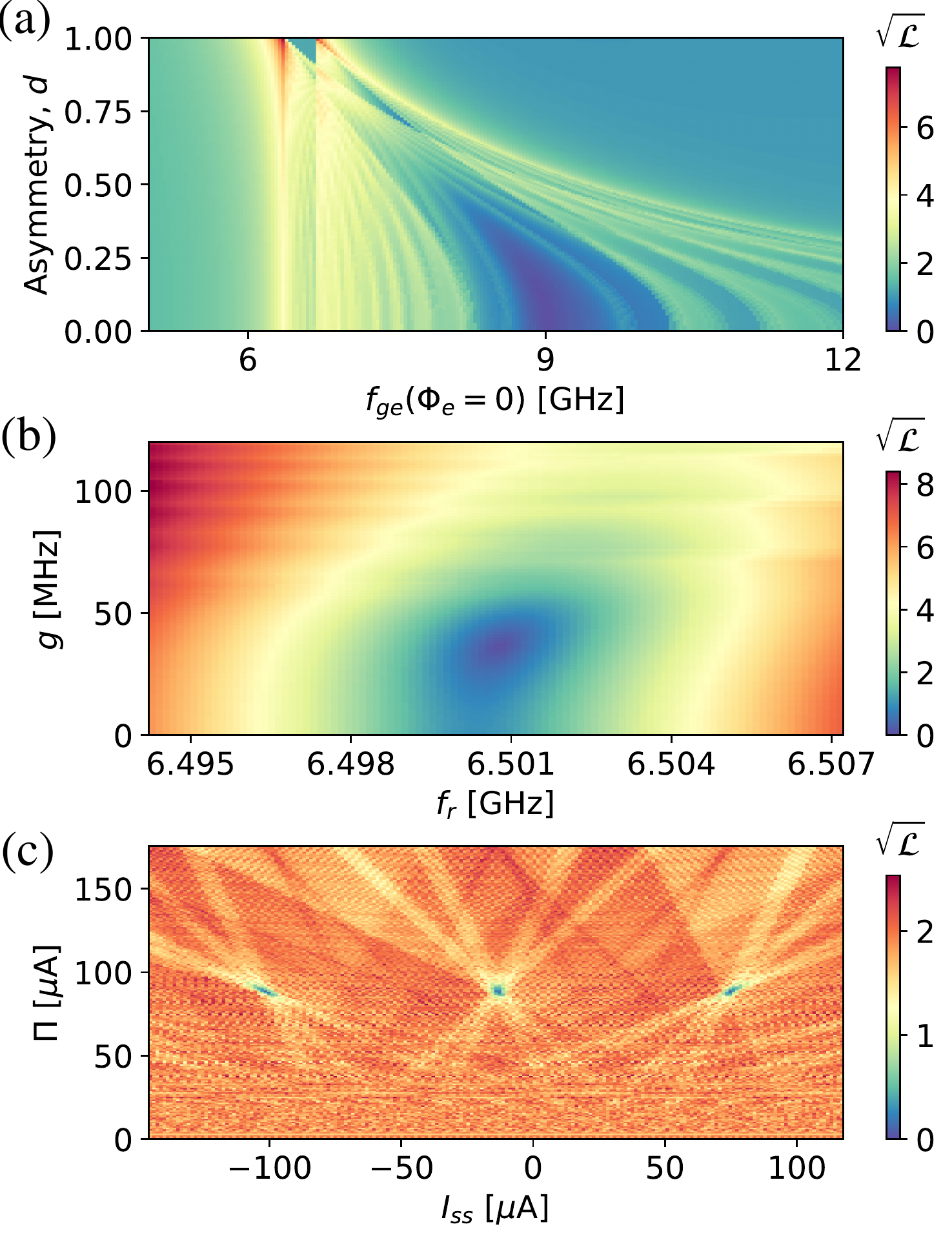}
	\caption{Slices of the loss function \eqref{eq:loss} for the full model and experimental data from the example made around the optimal point. Root-mean-square per-point loss $\sqrt{\mathcal{L}}$ is in MHz. (a) $f_{ge}^{max}$ and $d$ are varied, other parameters optimal. A large valley is located near 9 GHz, and some smaller locally minimal ones are present all around. (b) $g$ and $f_r$ are varied, others optimal. The loss function for these two parameters is well-conditioned near the optimum. (c) $\Pi$ and $I_{ss}$ varied, others optimal. This subplot illustrates a complex structure of local minima around the true one which we find analytically.}
	\label{fig:loss}
\end{figure}

Having performed the aforementioned preliminary steps, it is now possible to fit the full model to the extracted spectral points. To do this, we employ brute force optimization combined with the Nelder-Mead simplex downhill algorithm\cite{nelder1965} to polish the brute force result. For both methods, we use a common loss function that is based on the maximum likelihood method (again, we presume Gaussian distribution of the points around the model). 
For the known probe frequency span of the data $\Delta f_p$ (\autoref{fig:anti_exp}(a), whole y-axis) and the set of $N$ extracted points $\{p_i\} = \{(I_i, f_{r,i})\}$, we calculate the loss function as
\begin{align}
\mathcal{L} &= \sum_{i=1}^N [f_{r,i} - \mathcal{M}(I_i,\ \Pi, \ I_{ss},\ f_c,\ g,\ f_{ge}^{max},\ d)]^2,\label{eq:loss}\\
\mathcal{M} &= \begin{cases}
f_+,\  |f_+ - f_c|< \Delta f_p/2 \\
f_-,\ \text{otherwise}, \label{eq:cond}
\end{cases}
\end{align}
where, as in the previous section, $$f_{\pm} = f_{\pm}(I_i,\ \Pi,\ I_{ss},\ \ f_c,\ g,\ f_{ge}^{max},\ d)$$
The condition \eqref{eq:cond} means that we choose only the model frequencies that lie within the window $\Delta f_p$ around the model $f_c$ parameter. This ensures, firstly, that in the optimum case we will not take any excess points outside the experimental frequency scan, and, secondly, that we have a singular model value for each current in case of avoided crossings.

To substantiate the choice of the optimization algorithms, we present in \autoref{fig:loss} three visualizations of the defined loss function. The plots show how the function behaves if a certain pair of 6 model parameters is varied while others are kept optimal. From the plots it is obvious that the loss function is ill-defined and has a lot of local minima. Moreover, it may not be smooth everywhere because of the condition \eqref{eq:cond}. The $f^{max}_{ge}$ and $d$ parameters present significant difficulty in terms of false minima and low curvature as can be seen from \autoref{fig:loss}(a). In contrast, $f_c$ presents the least difficulty, as can be seen from \autoref{fig:loss}(b). The last plot \autoref{fig:loss}(c) shows a very complex structure of the loss function and narrow optimal valleys and serves as an illustration of why the period and phase extraction algorithm is important.

\begin{table}
	\centering
	\begin{ruledtabular}
		\begin{tabular}{lll} 
			Parameter & Value range & Steps \\ 
			\hline
			$f_c$ & $\langle f_r \rangle_{I} \pm 1$ MHz & 3\\ 
			$g$ & 20 - 40 MHz & 5\\
			$f_{ge}^{max}$ &  4 - 12 GHz & 80 \\
			$d$& 0 - 0.9 & 9
		\end{tabular} 
	\end{ruledtabular}
	\caption{Grid specifications for the brute force algorithm for STS detection.}
	\label{tab:grid}
\end{table}

The brute force algorithm acts on the grid specified in \autoref{tab:grid}. The ranges in the grid are based on the usual design parameters of the qubit samples in our database, and the number of steps is chosen so that the algorithm reliably finds the optimal valley. After the coarse brute force optimization is done and the optimal valley is located, we apply the Nelder-Mead search on all 6 parameters.

\section{Results}

\begin{table*}
	\centering
	\begin{ruledtabular}
		\renewcommand{\arraystretch}{1.2}

		\begin{tabular}{*{13}{l}} 
			\multirow{2}{*}{\textbf{Parameter}} & 
			\multicolumn{4}{l}{\textbf{(a)}} & 
			\multicolumn{4}{l}{\textbf{(b)}} & \multicolumn{4}{l}{\textbf{(c)}}\\
			& Brute & N-M & $\min \sigma$ & TTS  & Brute & N-M & $\min \sigma$ & TTS  & Brute& N-M & $\min \sigma$ & TTS  \\
			\hline
			$f_c$, GHz &6.5003 & 6.5007 &  4$\times 10^{-6}$ & n/a & 6.962 & 6.9631 & 1.6 $\times 10^{-4}$  & n/a &  6.47 & 6.465& 2$\times 10^{-4}$ & n/a\\ 
			$g$, MHz & 24 & 35.8 & 0.17 & n/a & 36 & 64.9 & 15 & n/a & 36 & 86.1& 1 &n/a\\
			$f_{ge}^{max}$, GHz & 8 &\textbf{8.97} & 0.013 & \textbf{9.04} &10.3& \textbf{11.3}& 1.35 & \textbf{9.08}& 6.3& \textbf{5.89}&0.01&\textbf{5.9}\\
			$d$ &0.5& \textbf{0.09}& 0.013& \textbf{0.13} &0.5&\textbf{0.49} &0.07&\textbf{0.6}&0.1& \textbf{0.25} & 0.05 &\textbf{0.3} \\\hline
			Loss, kHz & 251 & 20 && &338& 51 & & &2038& 149&&
		\end{tabular} 
	\end{ruledtabular}
	\caption{Analysis of optimal parameters found for the three cases in \autoref{fig:anti_fit_cases} after the brute and Nelder-Mead (N-M) optimizations. $\min \sigma$ columns stand for the lower bounds of the standard deviations of the MLE estimators found using the Fisher information matrix at the optimum.}
	\label{tab:sts_results}
\end{table*}

\begin{figure*}
	\centering
	\includegraphics[width=\linewidth]{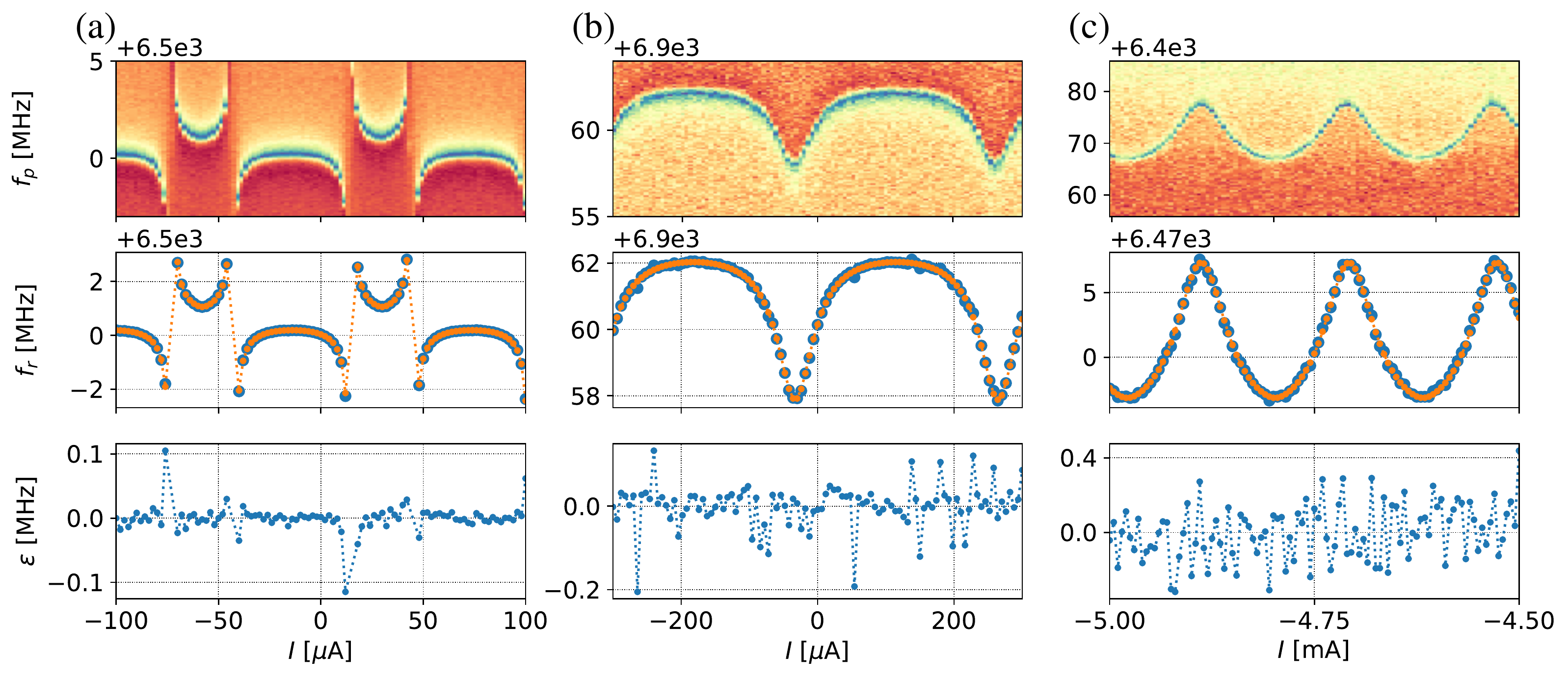}
	\caption{The results of the algorithm execution on three experimental examples from our database. In the upper row, original images are presented. In the middle row, the data  (blue dots) and the fitted model (orange connected dots) are shown. In the lower row are errors $\varepsilon$. (a) Avoided crossings pattern; per-point RMS error around 30 kHz. (b) The qubit entirely above the resonator; per-point RMS error around 60 kHz. (c) The qubit below the resonator; per-point RMS error around 150 kHz.}
	\label{fig:anti_fit_cases}
\end{figure*}

The above described algorithm was implemented in Python using  routines of the SciPy\cite{scipy} library. We present here examples of the detection for all possible qubit-resonator dispositions.  The resulting fits are shown in \autoref{fig:anti_fit_cases}(a)-(c) along with the original data and the error plots characterizing the quality of fit quantitatively; we will further reference them as (a), (b) and (c) respectively.

\subsection{Fit quality analysis} 

In \autoref{tab:sts_results} we analyse the fitting results for the three cases shown in \autoref{fig:anti_fit_cases}(a)-(c). Comparing the brute estimation and the final result after the Nelder-Mead search, one can see that the chosen algorithm order works as expected: the exhaustive search finds good initial conditions for the Nelder-Mead, and there is a significant improvement in the loss value of the polished result compared to the brute estimation which is due to the more accurate determination of the cavity frequency $f_c$ which leads as well to major shifts in optimal $g$, $f_{ge}^{max}$ and $d$.

However, in turns out that the optimal qubit parameters and the coupling strength may have some errors. For example, in case (b) for $f_{ge}^{max}$ we see a more than 2 GHz difference between the optimal and the more accurate value found for the sample with TTS. For (a) and (c), the differences in $f_{ge}^{max}$ are much smaller (about 10-70 MHz) but instead we see higher errors in $d$ (around 10\%). Inaccuracy in the qubit parameters may be qualitatively explained by the low sensitivity of the resonator frequency to the qubit frequency when they are far away (see \eqref{eq:f_r}) and by the strong correlations between the parameters.

To quantify the inaccuracy, we have calculated analytically the Fisher information matrices $\hat I$ at the optimal points for the three cases using the \textit{SymPy}\cite{sympy} package (see Appendix \ref{sec:ML} for details). In \autoref{tab:sts_results}, columns $\min \sigma$ we show the square roots of the Cramér-Rao lower bounds (CRLB) for the variances of our MLE estimators. Let us consider the estimates for $g$ and $f_{ge}^{max}$  parameters. It can be seen, in case (b) $\min \sigma$ are higher than $\min \sigma$ for them in the cases (a) and (c). This is in accordance with the high observed discrepancy between the TTS and estimated $f_{ge}^{max}$. Additionally, using the principal component analysis of the Hessians of the log-likelihood, we find that the most unconstrained component eigenvector contains the parameters $f_c,\ g,\ f_{ge}^{max},\ d$, and the second one (being an order of magnitude smaller) consists of $\Pi$ and $I_{ss}$. To reduce the correlations, one may choose another parametrization of the model, i.e. replace $d$ and $f_{ge}^{max}$ by $ E_{JJ}^{max} $ and $ E_{JJ}^{min} $. However, the accuracy of the fit using the presented parametrization is sufficient for practical applications.

\begin{figure*}
	\centering
	\includegraphics[width=.96\linewidth]{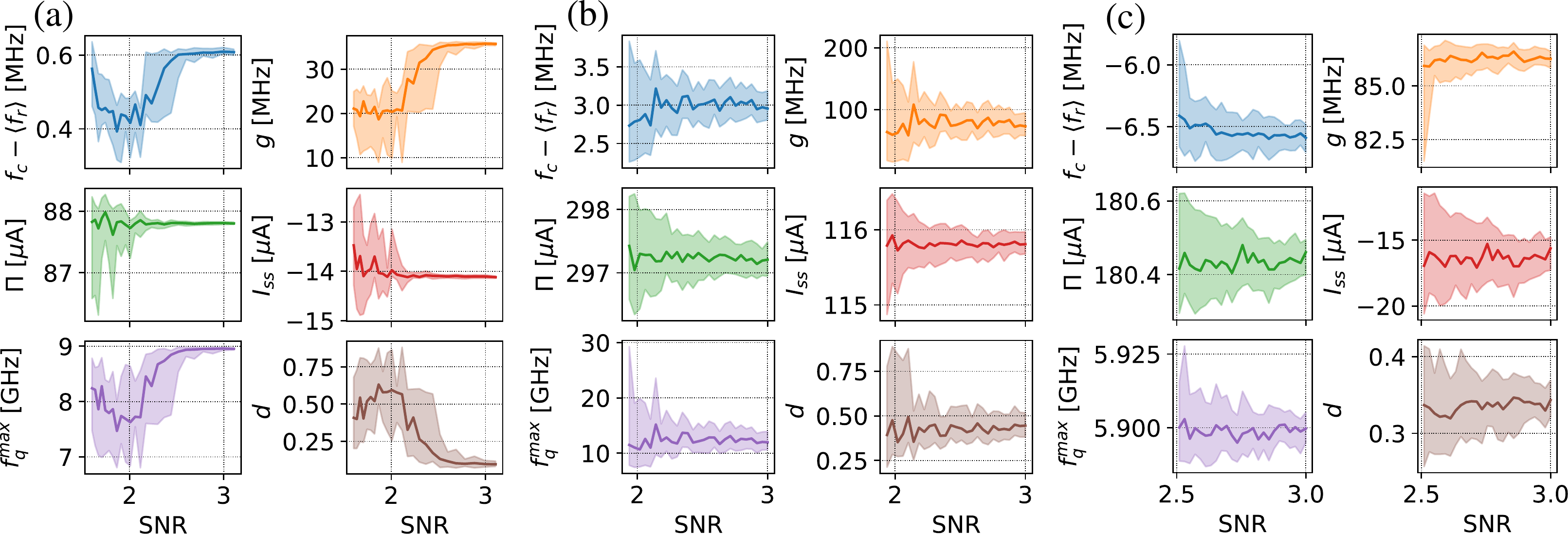}
	\caption{Behaviour of the algorithm on the real data from \autoref{fig:anti_fit_cases} with added gaussian noise of varying power (i.e., with artificially reduced SNR). The clouds show the 25th and 75th percentiles of the optimal parameter samples for 50 realizations of the added noise, solid lines show median values. (a) Original data SNR is 19. The algorithm is accurate above SNR=2.5 (well below the original SNR), but is still robust down to SNR=1.5. (b) Original data SNR is 4.7. The qubit frequency and asymmetry are not determined accurately, but the algorithm is stable above SNR=2. (c) Original data SNR is 3.14. The algorithm is robust above SNR=2.5, and is only noticeably uncertain in determining the asymmetry $d$.}
	\label{fig:noise_test}
\end{figure*}

\subsection{Testing noise robustness}

To stress test the algorithm, we artificially deteriorate the data from \autoref{fig:anti_fit_cases} by adding complex Gaussian noise of various variances to the complex data at each ($f_p$, $I$) before fitting. However, since the original pictures already contain an experimental noise, we first have to define a common scale for all three cases (a), (b) and (c) to be able to compare them. The signal-to-noise ratio (SNR) turns out to be a sensible universal measure of noise, and is defined for our data in a similar manner to the SNR in the resonator fitting tests\cite{probst2015}. Since the data for a resonator lie on a circle in the complex plane, it is convenient to take its radius $r$ as the signal amplitude. Then, if the total noise is of the form $(\xi_1+i\xi_2)/\sqrt 2$, where $\xi_1,\ \xi_2$ are distributed normally with zero mean and standard deviation (SD) $\sigma$, the SNR is defined as 
\begin{equation}
r/\sigma = r/\sqrt{\sigma_0^2+\sigma_1^2},
\label{eq:SNR}
\end{equation}
where $\sigma_0$ is the SD of the original noise and $\sigma_1$ is the SD of the added noise. Obviously, the resulting SNR in this scale can't be larger then the original SNR=$r/\sigma_0$. 

Using the described common scale for the measurements (a), (b) and (c), we test the algorithm stability by launching it 50 times for different added noise realizations with various standard deviations $\sigma_1$. The added noise power is steadily increased, and the resulting SNR is being calculated according to Eq. \eqref{eq:SNR}. We then record all fitting results, and calculate their mean value, 25\textsuperscript{th} and 75\textsuperscript{th} percentiles for each $\sigma_1$.  Since the original noise is fixed, the variance of the fitting results will go to zero at SNR=$ r/\sigma_0 $ where the added noise is negligible and the fit converges to the same values as in \autoref{tab:sts_results}. 

In \autoref{fig:noise_test} we plot the results of the test outlined above. As one can see from the graph, the algorithm becomes stable above SNR=3, and the lowest SNRs before divergence occurs are around 2. The original data SNR for (a) is 19, and thus in the tests $\sigma_1 > 5\sigma_0$. Conversely, the original data SNR for (b) and (c) are much lower, and thus $\sigma_1 \approx \sigma_0$ in the tests. This means that for (b) and (c) the variance of the fitting results shown with the clouds in \autoref{fig:noise_test} is lower than it would be if all the noise was generated.

In overall, the detection algorithm is stable even at very low SNRs where the patterns are barely visible by the human eye. The stability is limited mostly by the resonator frequency detection procedure.

\subsection{Performance}

\begin{table}[b]
	\begin{ruledtabular}
		\begin{tabular}{llllll}
			&Extract $f_r$& $\Pi$, $I_{ss}$ & Brute &N-M &
			\textbf{Total}\\\hline
			\textbf{Time}, s& 2.72& 0.3&2.79&1.37&7.34\\
			\textbf{Ratio}, \% & 37 &4 &38 &18 &100
		\end{tabular}
	\end{ruledtabular}
	\caption{Performance of the algorithm on a 2.5 GHz 2-core i5-5337U CPU for the case (a). Most of the time is taken by the $f_r$ extraction and the brute search steps.}
	\label{tab:performance}
\end{table}

We have tested the performance of the algorithm on a 5-year old 2-core Intel i5-3337U CPU @ 2.5 GHz. The results are presented in \autoref{tab:performance} for the case (a).  As can be seen, the procedure takes around 7 seconds to complete. On a contemporary Intel Core i7-7700 CPU, it takes about two seconds. 

The heaviest computational tasks are the frequency extraction part with the $circlefit$ library (2.7 seconds, most time spent in iterative calibration fits) and the brute search and Nelder-Mead optimization part where the time costs mostly come from the square root calculation necessary for Eqs. \eqref{eq:tr_levels} and \eqref{eq:branches2}. However, this is still fast enough, firstly, because recording a resonator spectrum such as one in the example takes around 20 seconds, and, secondly, it is much faster than doing it manually. 

Due to the complexity of calculating the value of the loss function, the frequency extraction scales at least as $\mathcal{O}(NM)$ if $N$ and $M$ are the number of points for current $I$ and probe frequency $f_p$, respectively, and the following optimization scales linearly with $N$. 

\section{Summary and discussion}

In summary, we have developed and implemented a method for automatic processing of single-tone spectroscopic data that uses maximum likelihood estimation to extract the physical parameters of cQED systems from measurement results. We have shown that it is possible to estimate all relevant circuit parameters by tracing only the behaviour of the resonance dip (or peak) for any kind of qubit-resonator disposition. Our method is very robust with respect to noise down to SNR=3 and is fast enough to be used in practice. The accuracy of the algorithm is not sufficient for starting qubit gate calibrations right after the single-tone spectroscopy step, but it facilitates much faster two-tone spectroscopy with significantly reduced scan ranges. Additionally, our automated evaluation method allows to distinguish between functional and defective samples.

The software implementation of the algorithm in Python can be found on GitHub\footnote{\url{https://github.com/vdrhtc/Measurement-automation/blob/master/lib2/fulaut/AnticrossingOracle.py}}. The code as well accepts data as a Python dictionary with corresponding keys. Currently, only transmon qubits are supported; however, implementing this analysis for another types of qubits seems rather straightforward.

\begin{figure}[t]
	\centering
	\includegraphics[width=\linewidth]{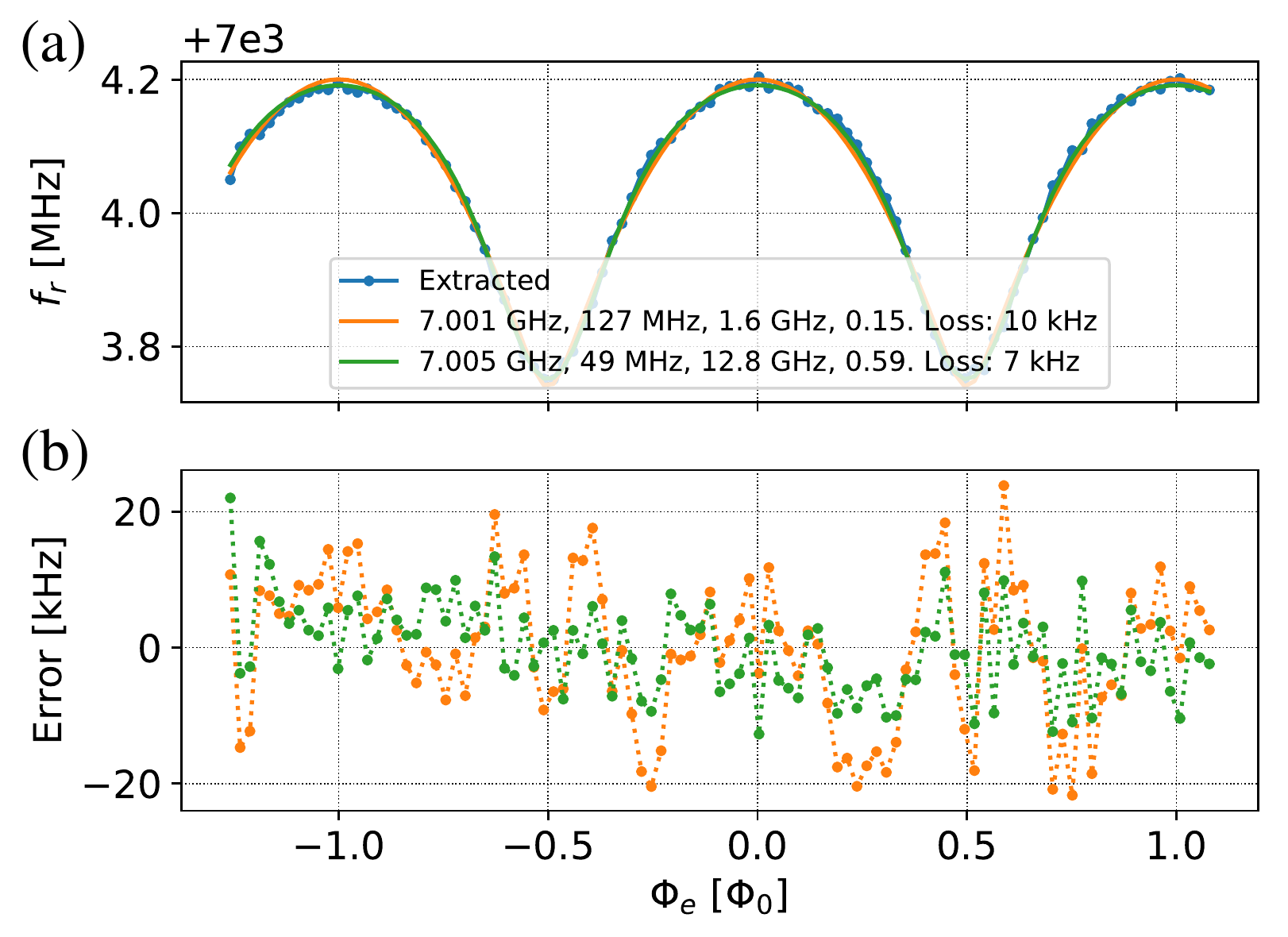}
	\caption{(a) Two alternative fits (orange and green lines) for the same data (blue dots). The corresponding parameters $f_c,\ g,\ f_{ge}^{max}, d$ for the two alternative fits are in shown the legend along with the loss values. (b) Residuals for the two fits, in orange (qubit below) and green (qubit above the resonator).}
	\label{fig:alternative-fits}
\end{figure}

Considering the downsides of the suggested approach, we note that the implementation of the algorithm is not perfectly optimized. The brute force search is not fully parallelized and might be replaced with a more sophisticated global optimization algorithm such as simulated annealing or basin hopping. Another problem that may slow down the execution is that for large qubit-cavity detuning and high noise two very different sets of parameters may minimize the loss function equally well. An illustration for this statement can be seen in \autoref{fig:alternative-fits}(a),(b) where the algorithm was launched twice using different grid ranges for $f_q^{max}$. In one case (orange) the search was done below, and in the other (green) above the mean resonance frequency $\langle f_c\rangle_{I}$. The resulting fits are almost equally accurate, and without additional information about the system it is not possible to reliably decide between the two. Therefore, some hints should be passed to the algorithm to avoid ambiguity; for instance, a prior probability density may be imposed on the parameter values and the MLE replaced with maximum a posteriori method. Otherwise, it would be necessary to check both possibilities with other methods, i.e. via two-tone spectroscopy. Since for the transmon qubits the case in \autoref{fig:alternative-fits} is the only type of ambiguity that may occur, we introduce an optional flag for the program that specifies where to look for the qubit transition frequency during the brute force search, above or below the frequency of the resonator.

In the future we are planning to use the proposed algorithm along with automated analysis of two-tone spectroscopic measurements to build a full system that will automatically calibrate the qubit samples from scratch.

\section{Acknowledgements}

We gratefully acknowledge valuable discussions with I. Besedin, V. Ryazanov, A. Dmitriev and P. Fedorov. 
This work was partially supported by the Ministry of Education and Science of Russian Federation in the framework of Increase Competitiveness Program of the NUST MISIS (contract no. K2-2017-081). Experiments and analytical tools were developed with the financial support from the Russian Science Foundation (contract no. 16-12-00095). Numerical simulations were performed with the support of Russian Foundation for Basic Research, grant No. 19-12-80006.

\appendix

\section{Transmon Hamiltonian}\label{sec:transmon}

\begin{figure}[t]
	\centering
	\includegraphics[width=\linewidth]{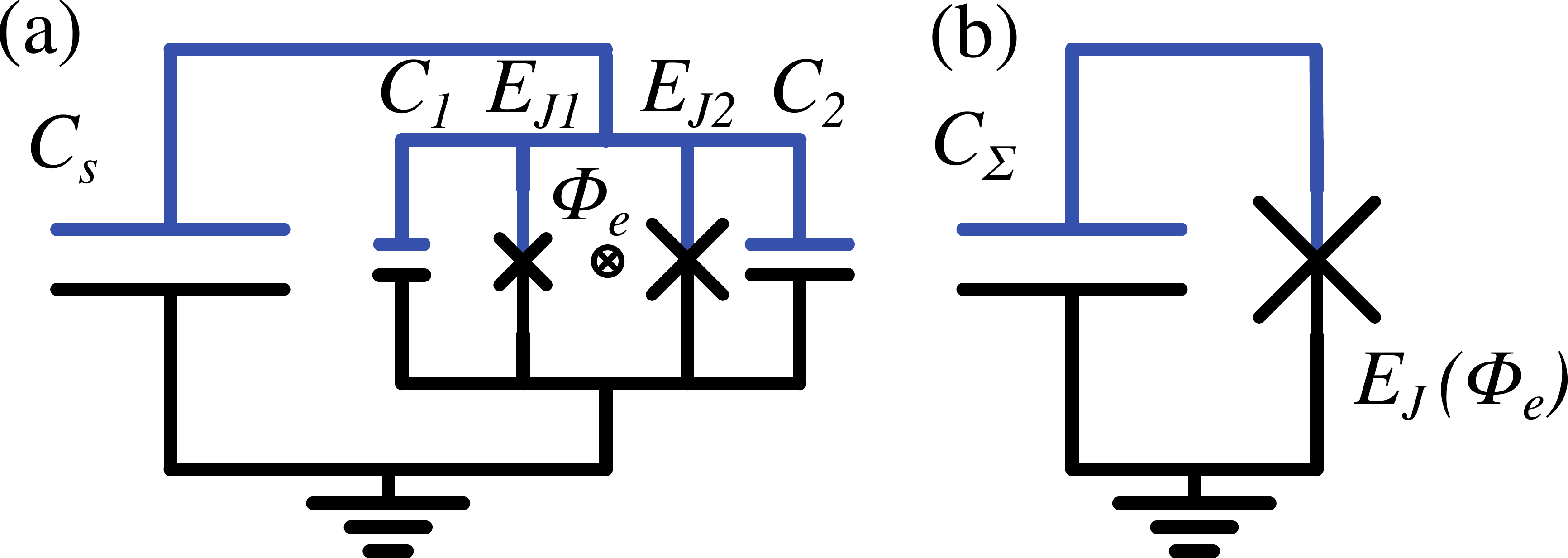}
	\caption{(a) A tunable transmon circuit with an asymmetric SQUID, $E_{J1} \neq E_{J2}$. (b) Equivalent transmon with tunable energy $E_{J}(\Phi_e)$ and unified capacitance $C_{\Sigma}$. The qubit island containing its single degrees of freedom is in blue.}
	\label{fig:trans}
\end{figure}

The simplest version of this qubit consists of a Josephson junction shunted with a large capacitor. Flux tunability of the frequency is attained by replacing a single Josephson junction with a SQUID as in \autoref{fig:trans}(a) and applying external magnetic flux $\Phi_e$ to its loop. This scheme can be equivalently represented by a capacitively shunted junction of tunable Josephson energy, \autoref{fig:trans}(b). The Hamiltonian for such system is as follows: 
\begin{equation}
\hat{H}_{tr} = 4E_C \hat n^2 - E_J(\Phi_e) \cos \hat\varphi,
\label{eq:tr_ham}
\end{equation}
where $E_C = e^2/2C_{\Sigma}$, $C_{\Sigma} = C_s + C_1 +C_2$, is the charging energy, $\hat n$ and $\hat \varphi$ are the operators for the Cooper pair number and the phase of the qubit island. For the equivalent Josephson energy $E_{J}(\Phi_e)$ one obtains
\begin{equation}
\begin{aligned}
E_{J}(\Phi_e) &= E_{J\Sigma} \cdot k(d, \Phi_e),\\
k(d, \Phi_e) &= \left[\cos^2\left(\frac{\pi \Phi_e}{\Phi_0}\right) +d^2 \sin^2 \left(\frac{\pi \Phi_e}{\Phi_0}\right)\right]^{\frac{1}{2}}
\label{eq:EJ_Phie}
\end{aligned}
\end{equation}  
where $E_{J\Sigma} = E_{J1}+E_{J2}$ ($E_{J1},\ E_{J2}$ are the single junction energies), $d = \frac{E_{J1}-E_{J2}}{E_{J1}+E_{J2}}$ is the asymmetry of the SQUID. As one can notice, the dependence is periodic in $\Phi_e$.

It is also possible to derive analytical expressions for the energy levels and transition frequencies for this type of qubits. The energy of the $m$\textsuperscript{th} level is (in the limit of $E_J\gg E_C$) \cite{koch2007}
\begin{equation}
E_m = m \sqrt{8E_J(\Phi_e) E_C} -\frac{E_C}{12}(6m^2+6m+3).
\label{eq:tr_levels}
\end{equation}
From this equation, the qubit transition frequency ($ \ket{0}\rightarrow \ket{1}$, or $ \ket{g}\rightarrow \ket{e}$, or the $ge$ transition) may be approximated as 
\begin{equation}
f_{ge}(\Phi_e) \approx  f_{ge}^{max} \cdot \sqrt{k(d, \Phi_e)}
\label{eq:tr_model}
\end{equation}
where $f_{ge}^{max} = \sqrt{8 E_J(0) E_C}-E_C$ is the maximal possible frequency. The error of this approximation is 
$$
\left[\sqrt{k(d, \Phi_e)}-1\right] \cdot E_C
$$  
which is non-negligible when $d$ is small and $\Phi_e\approx \Phi_0/2 $. This is satisfactory since the expression \eqref{eq:tr_levels} itself is no longer valid in these conditions because $E_J \approx E_C$ there. On the contrary, this approximation simplifies the model since now it depends just on two parameters ($f_{ge}^{max}, d$) instead of three. Therefore, we choose \eqref{eq:tr_model} as the theoretical curve in fitting.

One final note is that in the real-life applications is not possible to know directly the flux $\Phi_e$ that is threaded through the SQUID. The experimenter usually knows only the current $I$ (or voltage) which he applies to some coil that is connected inductively to the SQUID. Then $\Phi_e = M I + \Phi_r$, where $M$ stands for the mutual inductance of the coil and the SQUID, and $\Phi_r$ is some residual flux inherent to the sample. So, in the main text we only use the current $I$, the period in current $\Pi = \Phi_0/M$ and the sweet spot current $I_{ss} = -\Phi_r/M$ and not the corresponding magnetic fluxes.

\section{Circuit QED}\label{sec:cqed}

\begin{figure*}
	\centering
	\includegraphics[width=\textwidth]{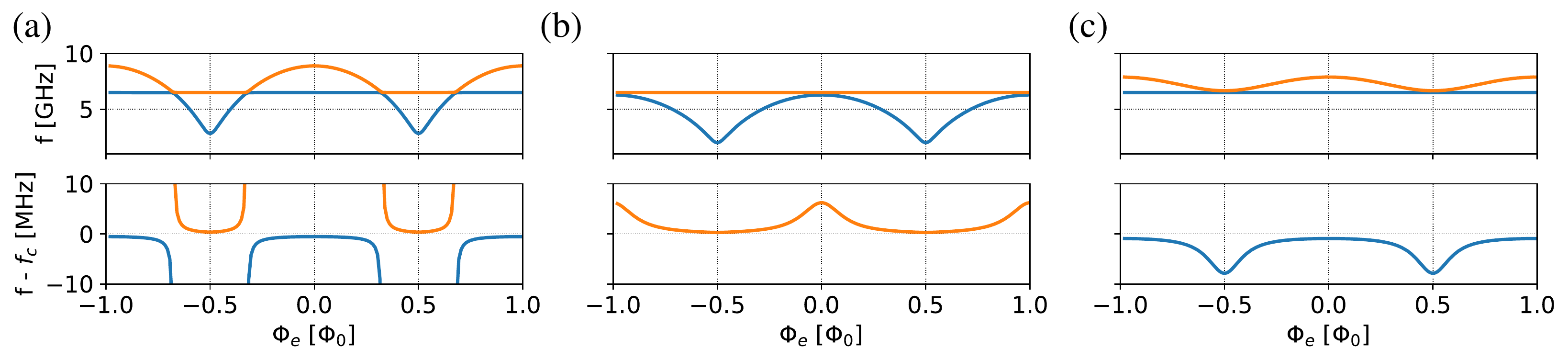}
	\caption{Frequency spectrum of the transmon-resonator system. Parameters used: $f_{ge}^{max} \approx \sqrt{8E_C E_J(0)}/2\pi = 8.5$ GHz, $d=0.3$, $f_r=6.4$ GHz, $g = 30$ MHz. For each subplot two transition branches  $f_{\pm} = (E_{\pm,0} - E_{g,0})/2\pi$ are shown (orange and blue, respectively) both forming the resonator and qubit lines. As one can notice, there are three qualitatively different cases of the resonator-qubit disposition. Lower row shows a zoomed area around $f_c$ that looks differently in each case.}
	\label{fig:anti_theor}
\end{figure*}

The readout of the superconducting qubits is now predominantly done using an ancilla system which is usually implemented as a superconducting microwave resonator which acts as an electromagnetic cavity in the standard cavity quantum electrodynamics (QED). Truncating the qubit to two levels ($ge$ transition), one may obtain the following Hamiltonian for the compound cavity-qubit system (in RWA):
\begin{equation}
\hat H/h = \frac{f_{ge}}{2} \hat \sigma_z + f_c \hat a^\dagger \hat a + g(\hat \sigma^- \hat a^\dagger + \hat \sigma^+ \hat a),
\end{equation}
where $f_{ge}$ is the qubit frequency, $f_c$ is the cavity frequency and $g$ is the coupling strength. 

As long as the RWA is done, this Hamiltonian may be diagonalized analytically\cite{blais2004}:
\begin{align}
E_{g, 0}/h &= \frac{f_c - f_{ge}}{2},\label{eq:branches1}
\\
E_{\pm, n}/h &= (n+1)f_c \pm \frac{1}{2}\sqrt{4g^2(n+1)+(f_{ge}-f_c)^2}.
\label{eq:branches2}
\end{align}

By substituting the dependence of the qubit frequency $f_{ge} \equiv f_{ge}(\Phi_e)$ into these equations, we can get straightforwardly the full system spectrum in dependence on the magnetic flux. We note here that, in general, $g$ may as well not be constant since the relevant matrix elements of the qubit may depend on the control parameters just as its frequency. For the transmon\cite{koch2007}
\[
g(\Phi_e) \propto |\bra{g}\hat n \ket{e}| \approx \frac{1}{\sqrt{2}}\left(\frac{E_J(\Phi_e)}{8 E_C}\right)^{1/4}.\label{eq:g}
\]
There are two useful observations from the expressions \eqref{eq:branches2} and \eqref{eq:g}. The first one is that for the avoided crossings the exact value of $g$ is only relevant when $(f_{ge} - f_c)^2/4 \lesssim g^2$. Therefore, if its dependence on the control parameters is slow enough (like for the transmon case), a constant $g$ fixed to its value $g(\Phi_e^*),\ \Phi_e^*: f_{ge}(\Phi_e^*) = f_c$ is a good approximation in the case (a) of \autoref{fig:anti_theor}. For the cases (b) and (c), in contrast, the constant $g$ model is, in general, not valid. Fortunately, for the transmon $g^2(\Phi_e) = g_{max}^2 \sqrt{k(d, \Phi_e)} \propto f_{ge}(\Phi_e)$ and thus equations \eqref{eq:branches2} may be rewritten in terms of $f_c' = f_c - g_{max}^2/f_{ge}^{max}$ and $g' = g_{max}\cdot \sqrt{f_c/f_{ge}^{max}}$ (note that now $g'$ is flux-independent). These approximations are accurate for \autoref{fig:anti_theor} (b), (c) when $g\ll f_c$. Our algorithm thus finds $f_c'$ and $g'$ while $f_{ge}^{max}$ and $d$ are not affected. For other types of qubits, in the linear coupling regime $g$ should be proportional to the matrix element of the qubit-resonator coupling operator. Therefore, the model for the algorithm needs to include $g(\Phi_e) = g' \cdot \bra{g}\hat V\ket{e}(\Phi_e) $ explicitly where $g'$ is the fitting parameter and $\hat V$ is the dimensionless qubit-resonator coupling operator term in the Hamiltonian.

In \autoref{fig:anti_theor} we have used the equations \eqref{eq:tr_model}, \eqref{eq:branches1} and \eqref{eq:branches2} to model a tunable transmon interacting with a cavity for various $\Phi_e$ and various $f_{ge}^{max},\ d$. In the lower row of the figure, one can see that it is possible to extract the dependence of the \textit{modified} cavity frequency $f_c^\prime$ on $\Phi_e$ (in the main text, we call it $f_r$ as the experimentally observed resonator frequency); for example, the avoided crossings pattern can be directly observed in \autoref{fig:anti_theor}(a), and the other two possible behaviours for the qubit entirely above or below the resonator in \autoref{fig:anti_theor}(b),(c). To shorten the notation, in the following we will define the corresponding branch frequencies of \eqref{eq:branches2} as $f_{\pm} = ( E_{\pm,0}-E_{g,0})/2\pi$.

\section{Maximum likelihood, Fisher information}\label{sec:ML}

For a vector $\mathbf{w}$ of parameters, we can calculate the likelihood $ \mathcal{L}(\mathbf{p}|\mathbf{w}) $ to observe $N$ sample data points $\mathbf{p} = \{I_i, f_{r,i}\}_{i=1}^N$ if they are presumed to be normally distributed around the model ($\mathbb{N}_{\mu_i, \sigma}$, $\mu_i = \mathcal{M}(I_i, \mathbf{w}),\ \sigma = \text{const}$) and uncorrelated. The sought value is the product of the probability densities for observing the individual data points:
\begin{equation}
\mathcal{L}(\mathbf{p}|\mathbf{w}) = \prod_{i=1}^{N} \frac{1}{\sqrt{2\pi}\sigma} \exp[ -(f_{r,i} - \mu_i)^2 / 2 \sigma^2]\label{eq:MLE} 
\end{equation}
The maximum likelihood method consists of finding the optimal parameter vector $\mathbf{w}^*$ such that $ \mathcal{L}(\mathbf{p}|\mathbf{w}^*) $ is the highest among all allowed $\mathbf{w} $. The vector $\mathbf{w}^*$ is called an estimator for the true parameter vector $\mathbf{w}^0$ and is itself a random value. If we take the negative logarithm of the likelihood, we will get the negative log-likelihood function:
\begin{align*}
- \ln \mathcal{L}(\mathbf{p}|\mathbf{w}) &= \sum_{i=1}^N (f_{r,i} - \mu_i)^2 / 2 \sigma^2 + N \ln\sqrt{2\pi}\sigma
\label{eq:logL} \\
&\equiv \chi^2/2 + N\ln \sqrt{2\pi}\sigma
\end{align*}
Minimizing this function is equivalent to maximizing the likelihood since the logarithm is monotonic. As the second term in the right part does not depend on $\mathbf{w}$, the task is reduced to minimizing the sum of squares, or the $\chi^2$; this argument motivates the choice of the form for the loss function \eqref{eq:loss}.

Next, we consider the variance of the MLE estimator $\mathbf w^*$ that we get from the fit. Neglecting its bias (in general, for nonlinear MLE it is always present but can only be calculated approximately\cite{cox1968} and scales as $1/N$), we can use the multivariate Cramér-Rao inequality to estimate the lower bound of its variance. In the general case, the inequality is as follows\cite{keener2011, schervish2012}:
\begin{equation}
\text{Var}_\mathbf{w}[\mathbf{w}^*] \geq \text{diag} [\hat I(\mathbf{w})^{-1}],\ \forall\,\mathbf{w}.
\label{eq:cramer-rao}
\end{equation} 
Here $\text{Var}_\mathbf{w}$ means the variance of the estimator when the \emph{true} model distribution parameters are $\mathbf{w}$, and $ \hat I(\mathbf{w}) $ is the Fisher information matrix. We show below that this matrix depends on $\mathbf{w}$, so to get the relevant bound we need to calculate it at $ \mathbf{w} = \mathbf{w}^0$. Since $\mathbf{w}^0$ is not known for the real data, one will have to use $\mathbf{w}^*$ as an approximation for $\mathbf{w}^0$. Using the definition of $\hat I(\mathbf{w})$, the properties of $\ln \mathcal{L}$ and assuming the possibility of double differentiation, we can write\cite{keener2011, schervish2012}:
\begin{align*}
\hat I(\mathbf{w}) 
&= E_\mathbf{w} \left[ \grad_\mathbf{w} \ln \mathcal{L} \cdot \grad_\mathbf{w}^T \ln \mathcal{L} \right]\\
&= E_\mathbf{w} \left[ - \mathbb{H}_\mathbf{w} \ln \mathcal{L} \right],\
\end{align*}
where $E_\mathbf{w}$ means the expectation under $\mathcal{L}(\mathbf{p}|\mathbf{w})$, i.e. the average over possible realizations of $\mathbf{p}$ distributed according to \eqref{eq:MLE}, and 
\[
\mathbb{H}_\mathbf{w} = 
\left(\begin{matrix}
\partial^2/\partial w_1^2 & \partial^2/\partial w_1 \partial w_2 & \dots \\
\partial^2/\partial w_1 \partial w_2
& \partial^2/\partial w_2^2 & \dots\\
\vdots & \vdots & \ddots
\end{matrix}\right) 
\]
is the Hessian operator. Next, using \eqref{eq:MLE} and the fact that $E_\mathbf{w}[f_{r,i}] = \mu_i(I_i, \mathbf{w})$, we can derive the analytic expression for $ \hat I(\mathbf{w}$):
\begin{equation}
\hat I(\mathbf{w}) = \sum_i \frac{\mathbb{H}_\mathbf{w} \mu_i^2(I_i, \mathbf{w}) - 2 \mu_i(I_i, \mathbf{w}) \mathbb{H}_\mathbf{w} \mu_i(I_i, \mathbf{w})}{2\sigma^2}.
\label{eq:fisher_analytic}
\end{equation}

Notably, this formula is valid for any type of nonlinear MLE with Gaussian noise. In overall, to obtain the Fisher information matrix, one needs, firstly, to calculate the Hessian of the model function $\mathcal{M}(I, \mathbf{w})$ at $\mathbf{w}=\mathbf{w}^*$ (or $ \mathbf{w}^0 $ if the data is synthetic). Secondly, the measurement error variance $\sigma^2$ is required. A common method to estimate it is to estimate the variance based on the sample $\{f_{r,i} - \mu_i\}$. However, for the nonlinear models it is not possible\cite{ye1998, andrae2010} to derive a valid analytic expression for the corresponding estimator $(\sigma^2)^*$ due to the unknown number of the effective degrees of freedom. One way to solve this problem is to find $\sigma^2$ independently by measuring the variance of $f_r$ at some fixed $I$. Another way is to use the expression valid for the linear regression
$(\sigma^{2})^* = \chi^2/(N-M)$ where $M = \dim \mathbf{w}$ and then test its validity using the Monte-Carlo approach. Our numerical tests with 5000 runs show that even when having $M=6$ and $N=18$, $(\sigma^{2})^*$ agrees well with the known $\sigma^2$ of the synthetic data:  $|E[(\sigma^{2})^*] - \sigma^2|< 0.01 \sigma^2$ and therefore can be used in error calculation.

\listoffixmes

\bibliography{papers_bibliography}
\end{document}